\documentclass[preprintnumbers,article,amsmath,amssymb,floatfix,10pt,prd,twocolumn,
superscriptaddress,nofootinbib]{revtex4-2}
\usepackage{bm}
\usepackage{amsfonts}
\usepackage{latexsym}
\usepackage[latin1]{inputenc}
\usepackage{graphicx}
\usepackage{amsmath}
\usepackage{palatino}
\usepackage{mathpazo}
\usepackage{textcomp}
\usepackage{float}
\usepackage{booktabs}
\usepackage{dcolumn}
\usepackage{ragged2e}
\usepackage{hyperref}
\usepackage{amsmath}
\usepackage{xcolor}
\usepackage{orcidlink}
\usepackage{epsfig}
\usepackage[caption=false]{subfig}
\usepackage{commath}
\usepackage{tabularx}
\usepackage{multirow}
\usepackage{mathtools}

\allowdisplaybreaks[1]

\begin{document}
	
	\color{black}       

	\title{Anisotropic Cosmology with interacting Dark Energy in $f(R,T)$ Gravity: A Data-Constrained \& independent Approach}
	
		\author{ S. H. Shekh\orcidlink{0000-0002-1932-8431}}\email{  da\_salim@rediff.com}
	\affiliation{L. N. Gumilyov Eurasian National University, Astana 010008, Kazakhstan.}
	\affiliation{Department of Mathematics, S.P.M. Science and Gilani Arts, Commerce College, Ghatanji, Yavatmal, \\Maharashtra-445301, India.}
	\affiliation{Pacif Institute of Cosmology and Selfology (PICS), Sagara, Sambalpur, Odisha 768224, India.}
	\author{ N. Myrzakulov\orcidlink{0000-0003}}\email{nmyrzakulov@gmail.com}
	\affiliation{L. N. Gumilyov Eurasian National University, Astana 010008, Kazakhstan.}

	\author{Anirudh Pradhan\orcidlink{0000-0002-1932-8431}}
	\email{pradhan.anirudh@gmail.com}
	\affiliation{Centre for Cosmology, Astrophysics and Space Science (CCASS), GLA University, Mathura-281406, U.P., India.}
	
	\author{M. Zeyauddin\orcidlink{0000-0001-8382-8994}}
	\email[]{uddin\_m@rcjy.edu.sa}
	\affiliation{Department of General Studies (Mathematics) Jubail Industrial College, Jubail 31961, Saudi Arabia.}
	\begin{abstract}
		\textbf{Abstract:} 
			In this work, we investigate the cosmological dynamics of an anisotropic Universe within the framework of $f(R,T)$ gravity by incorporating pressureless dark matter and the dark energy models. The analysis is carried out in a Bianchi type-I space-time, allowing us to capture possible deviations from isotropy and their evolution during cosmic expansion. A phenomenological reconstruction scheme based on a variable deceleration parameter is adopted to derive a redshift-dependent Hubble function. To establish observational viability, we constrain the free parameters of the model using a comprehensive statistical analysis that combines observational Hubble data and the Pantheon+ Type Ia supernova compilation. The resulting parameter space is tightly bounded, and the reconstructed expansion history exhibits strong consistency with current observational expectations. The model successfully reproduces the transition from an early decelerating phase to the present accelerated epoch, while asymptotically approaching a de Sitter-like regime.
			
			Further we analyzed the geometrical diagnostics, including the statefinder and $O_m$ diagnostics, which indicate a close correspondence with the standard $\Lambda$CDM scenario at late times. The behavior of the effective equation of state suggests a dynamically evolving dark energy component consistent with a quintessence-like regime. Additionally, the analysis of energy conditions confirms the physical admissibility of the model, whereas the stability investigation reveals the presence of classical instabilities at the perturbative level.
	
	\end{abstract}

	\maketitle
	\section{Introduction}\label{I}
An observational advances over the past two decades have firmly established that the present Universe is undergoing the phase of accelerated expansion. Independent probe such as Type Ia supernovae \cite{Riess1998,Perlmutter1999}, cosmic microwave background (CMB) anisotropies \cite{Spergel2003,Planck2018}, baryon acoustic oscillations (BAO) \cite{Eisenstein2005}, and large-scale structure surveys consistently support this paradigm. Within the framework of general relativity (GR), this late-time acceleration is attributed to an exotic cosmic component known as dark energy (DE), which contributes nearly $70\%$ of the total energy density of the Universe.

The simplest candidate for the dark energy is the cosmological constant $\Lambda$. Despite its remarkable observational successes, the $\Lambda$CDM scenario suffer from several theoretical issues, notably the fine-tuning and a cosmic coincidence problems \cite{Weinberg1989,Carroll2001}. These shortcomings have been motivated the exploration of alternative explanations for cosmic acceleration, broadly classified into two categories: dynamical dark energy model and modified theories of gravity. Dynamical dark energy models introduce additional degrees of freedom in the matter sectors, such as scalar fields in quintessence \cite{Ratra1988,Caldwell1998}, phantom and quintom scenarios \cite{Caldwell2002,Guo2005}, and k-essence model \cite{Armendariz2001}. On the other hand, Modified theories of gravity have emerged as compelling alternative to explain the late-time accelerated expansion of the Universe without invoking an unknown dark energy component. The Modified gravity theories seek to explain a late-time acceleration by extending the geometric sector of Einstein's field equations. Prominent examples include $f(R)$ gravity \cite{Nojiri2006,Capozziello2006}, $f(G)$ gravity \cite{Nojiri2005}, and $f(R,T)$ gravity \cite{Harko2011}.

Among these, $f(R,T)$ gravity in which $T$ be the trace of the energy-momentum tensor originally proposed by Harko et al.~\cite{Harko2011}. He introduces a non-minimal coupling between matter and geometry through the trace $T$ of the energy momentum tensor. This coupling leads to significant deviations from General Relativity (GR), including the non-conservation of the energy momentum tensor, and the appearance of extra force terms acting on massive particles.
		In the recent years, considerable progress had been made in exploring cosmological implications to the $f(R,T)$ gravity. A major direction involve constructing viable cosmological model capable of describing both early- and late-time dynamics. In particular, anisotropic cosmological models, especially those based on Bianchi type spacetime, have been extensively investigated. Recent studies indicated that suitable choices of the functional form $f(R,T)=R+2f(T)$ can successfully reproduce accelerated expansion while the remaining consistent with thermodynamic laws~\cite{Singh2025}. The validity of the generalized second law of thermodynamics (GSLT) has also been examined in this context, suggested that the thermodynamic consistency as an important viability criterion.
		Another important avenue of research focuses on observational constraint, which is a central theme in modern cosmology. Recent works have incorporated multiple observational datasets, including Cosmic Chronometers, Baryon Acoustic Oscillations, and Type Ia Supernovae, to constrain model parameters. Multi-component fluid cosmological models have shown good agreement with observations, which indicates that modified gravity effects can effectively mimic dark energy behavior~\cite{Kumar2025}. These analyses typically employ Markov Chain Monte Carlo techniques, thereby establishing a direct connection between the theoretical predictions and an observational data.
	A notable trend in recent literature is the use of model-independent reconstruction technique. Gaussian Process-based approaches have been employed to reconstruct the functional form of $f(R,T)$ directly from observational datasets, reducing theoretical bias and provide a more robust understanding of cosmic evolution~\cite{Hernandez2024}. Such data-driven methods are increasingly becoming important in testing the viability of modified gravity theories.
		In addition to the cosmological dynamics, significant attention has been devoted to energy conditions and stability analysis in $f(R,T)$ gravity. Modified Raychaudhuri equations have been used to study the violation of classical energy conditions, particularly the null energy condition (NEC), which plays a crucial role in the realization of bouncing cosmologies and the avoidance of initial singularities~\cite{Sharif2025}. These studies suggest that the matter geometry coupling in $f(R,T)$ gravity can naturally lead to NEC violation without requiring exotic matter fields.
	Furthermore, several extensions of the standard $f(R,T)$ gravity framework have been proposed, including metric-affine formulation and higher-order generalizations. Metric-affine approach allow independent variations of the metric and connection, leading to the richer geometrical structures and novel cosmological solutions~\cite{Bahamonde2024}. Such extension enhanced the theoretical scope of modified gravity and provide new avenue for addressing fundamental cosmological problems.
		Apart from large-scale cosmology, $f(R,T)$ gravity has also been applied to various astrophysical phenomena. Recent investigations have explored wormhole solutions, compact stars, and black hole thermodynamics within this framework. It has been shown that the viable wormhole geometries constructed without invoking exotic matter due to effective contributions from modified gravity terms~\cite{Moraes2024}. Similarly, studies on compact objects demonstrate that the consistency with observational constraints from gravitational wave and X-ray data.
	Overall, recent developments in $f(R,T)$ gravity indicates a transition from purely theoretical toward observationally viable and data-driven cosmology. The integration of statistical methods, reconstruction techniques, and astrophysical observations has significantly improved the predictive capability of these models. However, the fundamental issue such as the origin of matter- geometry coupling and its implications for quantum gravity remain open and require further investigation. Also, these matter-geometry coupling introduces a novel cosmological dynamics and has been extensively explored in the recent years \cite{Shabani2014,Alvarenga2013}. An alternative and physically appealing approach to dark energy arises from the holographic principle, which originates from black hole thermodynamics and quantum gravity considerations \cite{tHooft1993,Susskind1995,Li2004,Tavayef2018,Moradpour2018,Jahromi2018}.
On the other hand, anisotropic cosmological model play an important role in understanding the early Universe and possible deviations from isotropy. Although the observations strongly support large-scale isotropy at present, small primordial anisotropies may have existed and subsequently decayed during cosmic evolution. Bianchi type-I space-time, being the simplest anisotropic generalization of the flat FLRW universe, provides a useful framework for investigating such effects \cite{Ellis1969,Collins1973}.

Motivated by these considerations, the present work investigates the cosmological dynamics of interacting dark energy models in an anisotropic Bianchi type-I universe within the framework of $f(R,T)$ gravity. We reconstruct the cosmological parameters in terms of redshift, analyze stability and energy conditions, and employ geometrical diagnostics to test observational viability. Our results demonstrate that the dark energy model provides unstable and physically consistent description of late-time cosmic acceleration while naturally approaching the $\Lambda$CDM limit.

	
	\section{Basic formalism of $f(R,T)$ gravity}\label{II}
	
	Modified theories of gravity provide a natural extension of Einstein's general relativity by allowing the gravitational Lagrangian to depend on additional geometric or matter quantities. Among these theories, $f(R,T)$ gravity has received considerable attention due to its explicit coupling between geometry and matter, which leads to novel cosmological implications. In this framework, the gravitational action is generalized to depend on both the Ricci scalar $R$ and the trace $T$ of the energy--momentum tensor.
		
	The action for $f(R,T)$ gravity is given by
	\begin{equation}
		S = \frac{1}{16\pi} \int f(R,T)\sqrt{-g}\, d^4x
		+ \int \mathcal{L}_m \sqrt{-g}\, d^4x,
		\label{1}
	\end{equation}
	where $g$ is the determinant of the metric tensor $g_{ij}$, $\mathcal{L}_m$ denotes the matter Lagrangian density, and natural units $G=c=1$ are used throughout this work.
	
	Varying the action \eqref{1} with respect to the metric tensor yields the modified gravitational field equations
	\begin{widetext}
	\begin{equation}
		f_R(R,T) R_{ij}
		- \frac{1}{2} f(R,T) g_{ij}
		+ \left(g_{ij}\Box - \nabla_i \nabla_j\right) f_R(R,T)
		= 8\pi T_{ij}
		- f_T(R,T) T_{ij}
		- f_T(R,T) \Theta_{ij},
		\label{2}
	\end{equation}
\end{widetext}
	where
	\[
	f_R(R,T) \equiv \frac{\partial f(R,T)}{\partial R}, \quad
	f_T(R,T) \equiv \frac{\partial f(R,T)}{\partial T},
	\]
	$\Box \equiv \nabla^k \nabla_k$ is the d'Alembertian operator, and $\nabla_i$ represents the covariant derivative associated with the Levi-Civita connection. The tensor $\Theta_{ij}$ is defined as
	\begin{equation}
		\Theta_{ij} \equiv g^{\alpha\beta}
		\frac{\delta T_{\alpha\beta}}{\delta g^{ij}},
		\label{3}
	\end{equation}
	which depends explicitly on the choice of matter Lagrangian $\mathcal{L}_m$.
		
	The energy--momentum tensor is defined by
	\begin{equation}
		T_{ij} = -\frac{2}{\sqrt{-g}}
		\frac{\delta (\sqrt{-g}\mathcal{L}_m)}{\delta g^{ij}}.
		\label{4}
	\end{equation}
	
	In cosmological applications, the cosmic fluid is often modeled as a perfect fluid, whose energy--momentum tensor takes the form
	\begin{equation}
		T_{ij} = (\rho + p) u_i u_j - p g_{ij},
		\label{5}
	\end{equation}
	where $\rho$ and $p$ denote the total energy density and isotropic pressure, respectively, and $u_i$ is the four velocity of the fluid satisfying $u_i u^i = 1$. Following standard practice in $f(R,T)$ cosmology, we choose the matter Lagrangian as
	\begin{equation}
		\mathcal{L}_m = -p,
		\label{6}
	\end{equation}
	which leads to a well-defined and physically meaningful energy--momentum tensor. With this choice, Eq.~\eqref{3} yields
	\begin{equation}
		\Theta_{ij} = -2 T_{ij} - p g_{ij}.
		\label{7}
	\end{equation}
In order to obtain exact cosmological solutions and maintain analytical tractability, we consider the simplest non-trivial functional form of $f(R,T)$ given by
	\begin{equation}
		f(R,T) = \lambda_1 R + \lambda_2 T,
		\label{8}
	\end{equation}
	where $\lambda_1$ and $\lambda_2$ are coupling constants. The case $\lambda_2 = 0$ reduces to standard general relativity, while $\lambda_2 \neq 0$ introduces a direct coupling between matter and geometry. For the choice \eqref{8}, the derivatives of $f(R,T)$ are $f_R = \lambda_1$ and $f_T = \lambda_2$, which simplifies the field equations considerably by eliminating higher-order derivative terms.
	
	A key feature of $f(R,T)$ gravity is the non-conservation of the energy--momentum tensor. Taking the covariant divergence of Eq.~\eqref{2} and using the Bianchi identities, one obtains
	\begin{equation}
		\nabla^i T_{ij}
		= \frac{f_T}{8\pi - f_T}
		\left[
		(T_{ij} + \Theta_{ij}) \nabla^i \ln f_T
		+ \nabla^i \Theta_{ij}
		- \frac{1}{2} g_{ij} \nabla^i T
		\right].
		\label{9}
	\end{equation}
	
	For the linear form \eqref{8}, where $f_T = \lambda_2 = \text{constant}$, Eq.~\eqref{9} reduces to
	\begin{equation}
		\nabla^i T_{ij}
		= \frac{\lambda_2}{8\pi + \lambda_2}
		\nabla_j p.
		\label{10}
	\end{equation}
	
	This relation implies that the energy--momentum tensor is generally not conserved unless $\lambda_2 = 0$ or the pressure gradient vanishes. Physically, this non-conservation can be interpreted as an effective exchange of energy between matter and the gravitational field, which may play a crucial role in driving late-time cosmic acceleration.
	
	
	
	
\section{Anisotropic space--time and field equations}

In order to explore possible deviations from isotropy in the early Universe and their subsequent decay at late times, we consider an anisotropic but spatially homogeneous Bianchi type-I space--time. This geometry represents the simplest generalization of the spatially flat FLRW universe and is well suited for studying anisotropic effects in cosmology. Hence, the Bianchi type-I line element is given by
\begin{equation}
	ds^2 = dt^2 - A^2(t)\,dx^2 - B^2(t)\,dy^2 - C^2(t)\,dz^2,
	\label{11}
\end{equation}
where $A(t)$, $B(t)$, and $C(t)$ are the directional scale factors along the $x$, $y$, and $z$ axes, respectively.

The average scale factor $a(t)$ is defined as
\begin{equation}
	a(t) = (ABC)^{1/3},
	\label{12}
\end{equation}
and the spatial volume of the Universe is
\begin{equation}
	V = a^3 = ABC.
	\label{13}
\end{equation}

The directional Hubble parameters are defined as
\begin{equation}
	H_x = \frac{\dot A}{A}, \quad
	H_y = \frac{\dot B}{B}, \quad
	H_z = \frac{\dot C}{C},
	\label{14}
\end{equation}
where an overdot denotes differentiation with respect to cosmic time $t$.

The mean Hubble parameter is then given by
\begin{equation}
	H = \frac{1}{3}(H_x + H_y + H_z)
	= \frac{\dot a}{a}.
	\label{15}
\end{equation}

The expansion scalar $\theta$ and the shear scalar $\sigma^2$ are defined as
\begin{equation}
	\theta = u^i_{\ ;i} = 3H,
	\label{16}
\end{equation}
\begin{equation}
	\sigma^2 = \frac{1}{2}
	\sum_{i=1}^{3} (H_i - H)^2,
	\label{17}
\end{equation}
where $H_i = \{H_x, H_y, H_z\}$.

The anisotropy parameter of the expansion is given by
\begin{equation}
	A_p = \frac{1}{3H^2} \sum_{i=1}^{3} (H_i - H)^2
	= \frac{2\sigma^2}{3H^2},
	\label{18}
\end{equation}
which provides a quantitative measure of the deviation from isotropic expansion.

The cosmic fluid is assumed to consist of two components such as \textit{Pressureless dark matter} (DM) whose energy momentum tensor is observed as
\begin{equation}
	T^{(m)\,i}_{\ \ \ \ j} = \text{diag}(\rho_m,0,0,0),
	\label{19}
\end{equation}
and the and \textit{the dark energy fluid} (DE) whose energy momentum tensor is taken as
\begin{equation}
	T^{(s)\,i}_{\ \ \ \ j}
	= \text{diag}(\rho_s,-p_s,-p_s,-p_s),
	\label{20}
\end{equation}
where where $\rho_m$ is the energy density of dark matter, $\rho_s$ and $p_s$ denote the energy density and pressure of the dark energy component, respectively.

The total energy--momentum tensor is therefore
\begin{equation}
	T^i_{\ j} = T^{(m)\,i}_{\ \ \ \ j} + T^{(s)\,i}_{\ \ \ \ j}.
	\label{21}
\end{equation}

Substituting Eqs.~\eqref{11} and \eqref{21} into Eq.~\eqref{2}, we obtain the modified Einstein field equations
\begin{equation}
	R_{ij} - \frac{1}{2} R g_{ij}
	= \frac{8\pi + \lambda_2}{\lambda_1} T_{ij}
	+ \frac{\lambda_2}{\lambda_1}
	\left(p + \frac{1}{2} T \right) g_{ij}.
	\label{22}
\end{equation}

where $p = p_s$ and $T = \rho_m + \rho_s - 3p_s$. Equation \eqref{22} clearly shows that the matter--geometry coupling effectively modifies the gravitational constant and introduces an additional source term proportional to the trace of the energy--momentum tensor. The independent components of the field equations are obtained as follows:
\begin{equation}
	\frac{\ddot B}{B}
	+ \frac{\ddot C}{C}
	+ \frac{\dot B \dot C}{BC}
	=
	\frac{16\pi + 3\lambda_2}{2\lambda_1} p_s
	- \frac{\lambda_2}{2\lambda_1}(\rho_m + \rho_s).
	\label{23}
\end{equation}
\begin{equation}
	\frac{\ddot C}{C}
	+ \frac{\ddot A}{A}
	+ \frac{\dot C \dot A}{CA}
	=
	\frac{16\pi + 3\lambda_2}{2\lambda_1} p_s
	- \frac{\lambda_2}{2\lambda_1}(\rho_m + \rho_s).
	\label{24}
\end{equation}
\begin{equation}
	\frac{\ddot A}{A}
	+ \frac{\ddot B}{B}
	+ \frac{\dot A \dot B}{AB}
	=
	\frac{16\pi + 3\lambda_2}{2\lambda_1} p_s
	- \frac{\lambda_2}{2\lambda_1}(\rho_m + \rho_s).
	\label{25}
\end{equation}
\begin{equation}
	\frac{\dot A \dot B}{AB}
	+ \frac{\dot B \dot C}{BC}
	+ \frac{\dot C \dot A}{CA}
	=
	-\,\frac{16\pi + 3\lambda_2}{2\lambda_1}
	(\rho_m + \rho_s)
	+ \frac{\lambda_2}{2\lambda_1} p_s.
	\label{26}
\end{equation}

Equations \eqref{23}--\eqref{26} constitute a system of four independent equations governing the dynamics of the anisotropic Universe in $f(R,T)$ gravity. It is therefore convenient to express the field equations directly in terms of the mean Hubble parameter $H$ and the total energy density.

The component of the field equations can be written as
\begin{equation}
	2\dot{H}+3H^2+\sigma^2
	=
	-\frac{16\pi+3\lambda_2}{2\lambda_1}p_s
	+\frac{\lambda_2}{2\lambda_1}(\rho_m+\rho_s).
\end{equation}
\begin{equation}
	3H^2-\sigma^2
	=
	-\frac{16\pi+3\lambda_2}{2\lambda_1}(\rho_m+\rho_s)
	+\frac{\lambda_2}{2\lambda_1}p_s,
\end{equation}
From the above field equations, we obtained $\rho_s$ and $p_s$ as
\begin{small}
\begin{equation}
	\rho_s
	=
	-\rho_m
	+
	\frac{2\lambda_1}{16\pi+3\lambda_2}
	\left(3H^2 - \sigma^2\right)
	+
	\frac{\lambda_2}{16\pi+3\lambda_2}
	\left(2\dot{H} + 3H^2 + \sigma^2\right).
	\label{29}
\end{equation}
\begin{equation}
	p_s
	=
	\frac{2\lambda_1}{16\pi+3\lambda_2}
	\left(2\dot{H} + 3H^2 + \sigma^2\right)
	+
	\frac{\lambda_2}{16\pi+3\lambda_2}
	\left(3H^2 - \sigma^2\right),
	\label{30}
\end{equation}
\end{small}

In order to analyze the anisotropic behavior of the cosmological model, the shear scalar $\sigma^2$ plays a fundamental role in quantifying the deviation from isotropy. In the present work, the field equations have been expressed in terms of the shear scalar with the help of Eqs. (\ref{17}) and (\ref{18}), which relate $\sigma^2$ and the anisotropy parameter to the directional Hubble rates. Since the expressions for the energy density $\rho_s$ and pressure $p_s$ explicitly depend on $\sigma^2$ [see Eqs. (\ref{29}) and (\ref{30})], it is essential to determine its functional form.

To obtain $\sigma^2$, we solve the system of field equations (\ref{23})-(\ref{26}), which govern the dynamics of the metric potentials $A(t)$, $B(t)$, and $C(t)$. Following the standard procedure in Bianchi type-I cosmology, the metric potentials are expressed in terms of the average scale factor $a(t)$ as $
A = a \exp\left(b_1 \int a^{-3} dt\right), \quad
B = a \exp\left(b_2 \int a^{-3} dt\right), \quad
C = a \exp\left(b_3 \int a^{-3} dt\right),
$
where $b_1$, $b_2$, and $b_3$ are constants satisfying the condition $b_1 + b_2 + b_3 = 0$, ensuring the consistency of the spatial volume $V = ABC = a^3$. Using these relations, the directional Hubble parameters are obtained as$
H_x = \frac{\dot{A}}{A} = H + \frac{b_1}{a^3}, \quad
H_y = \frac{\dot{B}}{B} = H + \frac{b_2}{a^3}, \quad
H_z = \frac{\dot{C}}{C} = H + \frac{b_3}{a^3}.
$
Substituting these expressions into Eq. (\ref{17}), the shear scalar takes the form $
\sigma^2 = \frac{1}{2} \sum_{i=1}^{3} (H_i - H)^2 = \frac{1}{2a^6} \left(b_1^2 + b_2^2 + b_3^2 \right).
$

Thus, the shear scalar is inversely proportional to the sixth power of the scale factor, which indicates that the anisotropy decays rapidly during the cosmic expansion. Defining a constant $\sigma_0^2 = \frac{1}{2}(b_1^2 + b_2^2 + b_3^2)$, the final expression for the shear scalar is obtained as
\begin{equation}
\sigma^2 = \frac{\sigma_0^2}{a^6}.
\end{equation}

\section{Solution with Redshift-Based Reconstruction}

Having established the expressions for the physical parameters in terms of the mean Hubble parameter and the shear scalar, we now proceed to determine the dynamical evolution of the model. Since the quantities such as $\rho_s$, $p_s$, and $\sigma^2$ are ultimately governed by the behavior of the Hubble parameter, it becomes essential to construct a suitable functional form of $H$ that can describe the expansion history of the Universe. Instead of solving the modified field equations directly for a specific matter content, we adopt a phenomenological approach by considering a variable deceleration parameter. This approach is well motivated in modern cosmology, as it allows for a flexible description of cosmic evolution, including the transition from an early decelerated phase to the present accelerated expansion. Moreover, it provides a convenient framework to express the Hubble parameter in terms of redshift, which is directly related to observations. The deceleration parameter is defined as
\begin{equation}\label{32}
	q = -1 - \frac{\dot{H}}{H^2},
\end{equation}
which provides a direct measure of the rate of change of the Hubble parameter and distinguishes between accelerating ($q<0$) and decelerating ($q>0$) phases of cosmic expansion. Using the relation 	$\frac{d}{dt} = -(1+z)H \frac{d}{dz}$, the time derivative of $H$ is obtained as 
\begin{equation}\label{33}
	\dot{H} = -(1+z)H \frac{dH}{dz}
\end{equation}
with the help of equation (\ref{33}), the equation (\ref{32}) becomes
\begin{equation}\label{34}
	q = -1 + (1+z)\frac{1}{H}\frac{dH}{dz}.
\end{equation}

In general, the deceleration parameter can be treated either as a function of cosmic time (or equivalently redshift), $q = q(t)$ or $q = q(z)$, or as a constant. A constant deceleration parameter provides a simplified yet physically insightful framework that leads to analytically tractable solutions of the field equations. This assumption corresponds to a power-law type expansion of the Universe and has been widely employed in cosmology to explore different evolutionary scenarios. In particular, a constant $q$ leads to a direct relationship between the Hubble parameter and redshift, making it suitable for phenomenological modeling and comparison with observations.\\
On the other hand, a variable deceleration parameter is more general and allows for a realistic description of cosmic evolution, including the transition from an early decelerating phase to the present accelerating phase. Such parametrizations have been extensively used in the literature to reconstruct the expansion history in a model-independent manner. Motivated by these considerations, here we consider the deceleration parameter dependence of the Hubble parameter as:
\begin{equation}\label{35}
	q =   \alpha + \frac{\beta}{H}.
\end{equation}
where $\alpha$ and $\beta$ are constants. 
From equation (\ref{34}) and (\ref{35}), we obtain
\begin{equation}
	-1 - \frac{\dot{H}}{H^2} = \alpha + \frac{\beta}{H}.
\end{equation}
Rearranging, we get the differential equation
\begin{equation}
	\dot{H} = -(1+\alpha)H^2 - \beta H.
\end{equation}
Solving this equation, we obtain
\begin{equation}
	H(t) = \frac{\beta}{C e^{\beta t} - (1+\alpha)}.
\end{equation}
Using the relation $H = \frac{\dot{a}}{a}$, the scale factor becomes
\begin{equation}
	a(t) = a_0 \left(C e^{\beta t} - (1+\alpha)\right)^{\frac{1}{1+\alpha}}.
\end{equation}
Now, using the definition of redshift
\begin{equation}
	1+z = \frac{a_0}{a(t)},
\end{equation}
we obtain the time and redshift relation as 
\begin{equation}
	t(z) = \frac{1}{\beta} 
	\ln \left[
	\frac{1+\alpha}{C \left(1 - (1+z)^{-(1+\alpha)}\right)}
	\right].
\end{equation}		
Thus, with this relation, the Hubble parameter becomes
\begin{equation}
	H(z) = H_0 \left[ 1 + \gamma \left((1+z)^{1+\alpha} - 1 \right) \right]
\end{equation}
where $H_0$ is the present value of the Hubble parameter and $\gamma = \frac{\beta}{H_0(1+\alpha)}$ be the dimensionless parameter. It is important to note that the anisotropic contribution encoded in the shear scalar $\sigma^2 \propto a^{-6}$ naturally diminishes at late times, ensuring that the model asymptotically approaches isotropy. Therefore, the late-time dynamics is effectively governed by the behavior of the reconstructed Hubble parameter, making the present approach consistent with observational expectations. In this context, we parametrize the deceleration parameter as a function of the Hubble parameter, which leads to a first-order differential equation for $H$. Solving this equation enables us to obtain explicit expressions for $H(t)$ and subsequently $H(z)$, which can be directly utilized for further physical and observational analysis.

\begin{center}
	\textbf{Data Sets}
\end{center}

A robust assessment of the cosmological viability of the proposed theoretical framework requires the precise estimation of its free parameters, namely $\alpha$, $\beta$, and $H_0$, through confrontation with observational data. In this regard, we employ a combination of Observational Hubble Data (OHD) and the Pantheon+ supernova compilation, both of which provide crucial insights into the expansion history of the Universe. While the OHD dataset offers direct measurements of the Hubble parameter $H(z)$, the Pantheon+ sample provides high-precision constraints through luminosity distance measurements of Type Ia supernovae.

\subsection{Observational $H(z)$ Data (OHD)}

To constrain the model parameters, we utilize a compilation of 77 uncorrelated $H(z)$ data points spanning the redshift range $0 \leq z \leq 2.36$. These measurements are obtained using different observational techniques, including the Cosmic Chronometer (CC) method, baryon acoustic oscillations (BAO) in galaxy clustering, and BAO measurements from the Ly$\alpha$ forest. The complete dataset, along with corresponding references, is summarized in Table \ref{tab}.

In order to incorporate systematic uncertainties associated with the CC measurements, we employ the full covariance matrix as provided by~\cite{Moresco2020}. The inclusion of the covariance matrix ensures a statistically consistent treatment of correlations among the data points, thereby enhancing the robustness of the parameter estimation.

The corresponding chi-square function is defined as
\begin{equation}
	\chi^2_{\mathrm{OHD}} = (\mathbf{H}_{\mathrm{obs}} - \mathbf{H}_{\mathrm{model}})^T \mathbf{C}^{-1} (\mathbf{H}_{\mathrm{obs}} - \mathbf{H}_{\mathrm{model}}),
\end{equation}
where $\mathbf{H}_{\mathrm{obs}}$ and $\mathbf{H}_{\mathrm{model}}$ denote the observed and theoretical values of the Hubble parameter, respectively, and $\mathbf{C}$ represents the covariance matrix.

\subsection{Pantheon+ Supernova Data}

To further constrain the cosmological parameters, we employ the latest Pantheon+ compilation of Type Ia supernovae \cite{Scolnic2018}, which represents one of the most comprehensive and precise datasets for probing the expansion history of the Universe. The sample consists of 1550 spectroscopically confirmed SNe Ia corresponding to 1701 light-curve measurements, covering the redshift range $z \in [0.001, 2.26]$. Due to their standardizable luminosity, SNe Ia serve as reliable distance indicators and play a crucial role in investigating late-time cosmic acceleration.

The theoretical luminosity distance is given by
\begin{equation}
	D_{L}(z) = (1+z)\int_{0}^{z}\frac{H_{0}}{H(z^{\prime})}\,dz^{\prime},
\end{equation}
from which the distance modulus is defined as
\begin{equation}
	\mu(z) = 5 \log_{10} \left(\frac{D_L(z)}{1~\mathrm{Mpc}}\right) + 25.
\end{equation}

It is important to note that the Pantheon+ dataset provides relative distance modulus measurements, which are not absolutely calibrated due to the degeneracy with the absolute magnitude $M_{B}$ of Type Ia supernovae. Therefore, the theoretical apparent magnitude is expressed as
\begin{equation}
	m^{\mathrm{th}}_{B,i} = \mu(z_i) + M_B,
\end{equation}
where $M_B$ acts as a nuisance parameter in the analysis.

The corresponding chi-square function is constructed as
\begin{equation}
	\chi^2_{\mathrm{SN}} = \Delta \mathbf{m}^T \, \mathcal{C}^{-1} \, \Delta \mathbf{m},
\end{equation}
where $\Delta \mathbf{m} = \mathbf{m}^{\mathrm{obs}} - \mathbf{m}^{\mathrm{th}}$, and $\mathcal{C}$ denotes the full covariance matrix associated with the Pantheon+ sample.

\medskip

\noindent
To estimate the model parameters, we employ a standard $\chi^2$ minimization procedure by comparing theoretical predictions with observational data. In general, the chi-square estimator is expressed as
\begin{equation}
	\chi^{2} = \sum_{i=1}^{N} \left[\frac{E_{\mathrm{th}}(z_i, \Phi_c) - E_{\mathrm{obs}}(z_i)}{\sigma_i}\right]^2,
\end{equation}
where $E_{\mathrm{th}}$ and $E_{\mathrm{obs}}$ represent the theoretical and observed quantities, respectively, $\sigma_i$ denotes the corresponding uncertainties, and $N$ is the total number of data points. The cosmological parameter set is defined as $\Phi_c = (H_0, \alpha, \beta)$, while $M_B$ is treated as a nuisance parameter in the supernova analysis.
\section{Results of observational constraints}

In order to examine the viability of the proposed cosmological model, we perform a comprehensive parameter estimation using three independent analyses: (i) Observational Hubble Data (OHD), (ii) Pantheon+ Supernova data, and (iii) their combined (joint) dataset. The free parameters of the model, namely $\gamma$, $\alpha$, and $H_0$, are constrained using a Markov Chain Monte Carlo (MCMC) approach by minimizing the corresponding $\chi^2$ functions.

\subsection{Constraints from OHD Data}

We first constrain the model parameters using the 77 uncorrelated observational Hubble data points spanning the redshift range $0 \leq z \leq 2.36$. The obtained best-fit values along with $1\sigma$ uncertainties 
with the minimum chi-square value are given in the Tab. \ref{tab1}.
 The relatively lower value of $H_0$ compared to standard cosmological estimates reflects the known tendency of OHD data to favor smaller Hubble constant values, primarily due to its direct and model-independent nature.

\subsection{Constraints from Pantheon+ Supernova Data}

Next, we utilize the Pantheon+ dataset consisting of 1701 data points to constrain the model parameters. The resulting best-fit with the minimum chi-square value are given in the Tab. \ref{tab1}. 
Compared to the OHD dataset, the Pantheon+ constraints exhibit larger uncertainties, particularly in $\alpha$, due to degeneracies between cosmological parameters and the nuisance parameter $M_B$. However, the inferred value of $H_0$ is consistent with standard cosmological measurements.

\subsection{Joint Constraints (OHD + Pantheon+)}

Finally, we perform a joint analysis by combining both OHD (77 points) and Pantheon+ (1701 points) datasets, thereby exploiting the complementary strengths of direct $H(z)$ measurements and high-precision luminosity distance observations.
The combined constraints are significantly tighter. The resulting best-fit with the minimum chi-square value are given in the Tab. \ref{tab1}. 
These results clearly demonstrate that the joint analysis effectively reduces parameter uncertainties and yields a value of $H_0$ that lies within the range reported by recent observations. The improvement in constraints highlights the importance of combining multiple datasets to break parameter degeneracies.

\subsection{Statistical Contours and Model Comparison}

\begin{figure}
	\centering
	\includegraphics[scale=0.60]{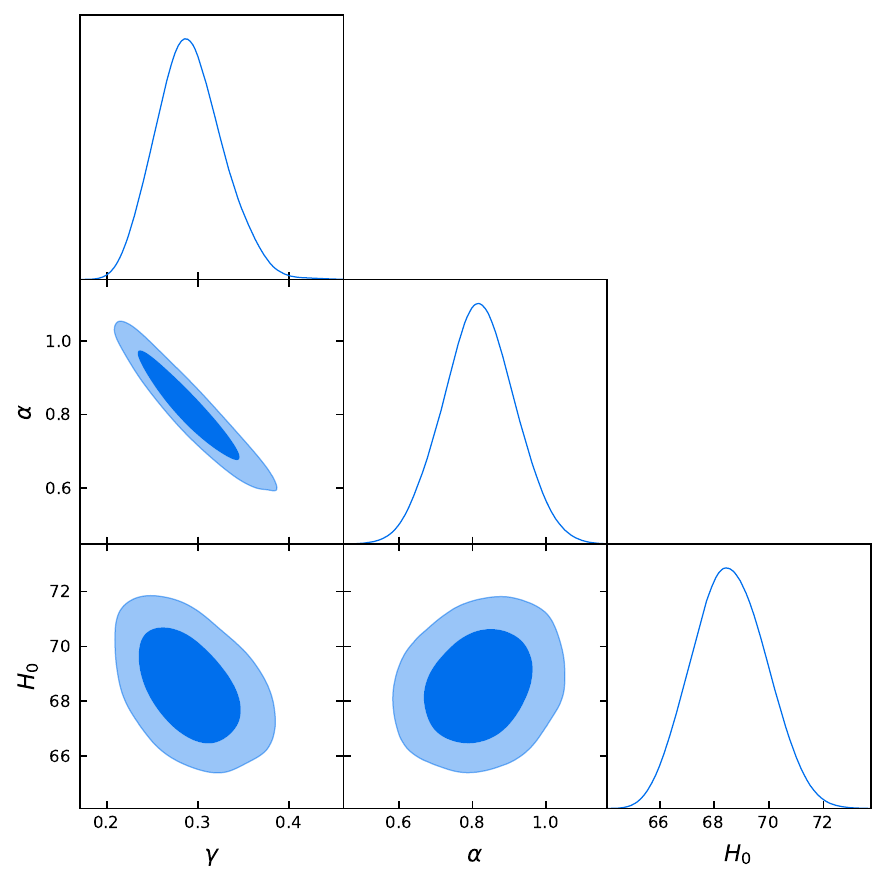}
	\caption{One-dimensional marginalized distributions and two-dimensional confidence contours at $1\sigma$ and $2\sigma$ levels for the joint analysis.}
	\label{joint_contours}
\end{figure}

The one-dimensional marginalized distributions and two-dimensional confidence contours at $1\sigma$ and $2\sigma$ levels for the model parameters are shown in Fig.~\ref{joint_contours}, obtained from the joint analysis. These contours illustrate the allowed parameter space and correlations among $\gamma$, $\alpha$, and $H_0$.

\begin{figure*}[ht]
	\centering
	\includegraphics[scale=0.50]{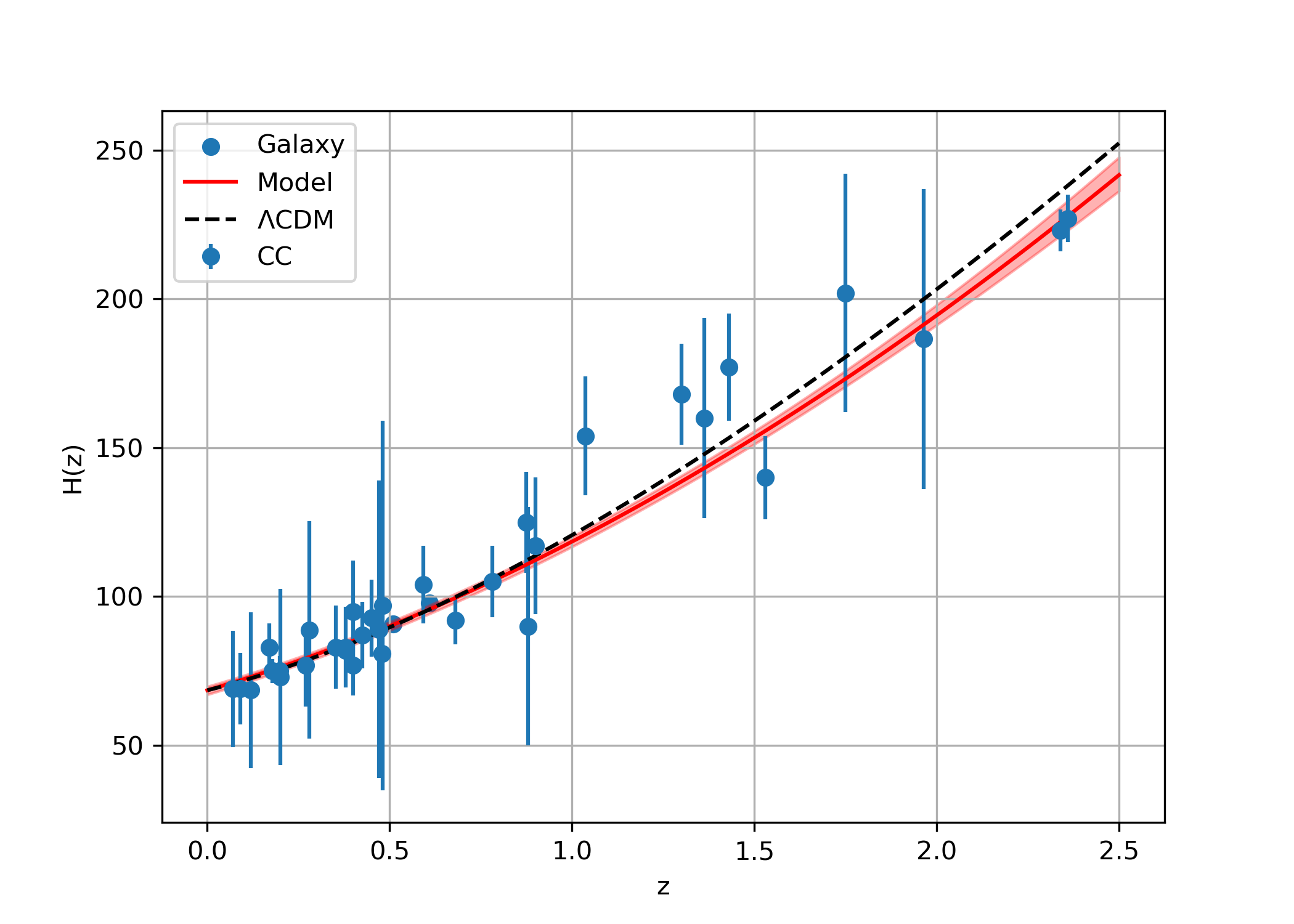}
	\includegraphics[scale=0.50]{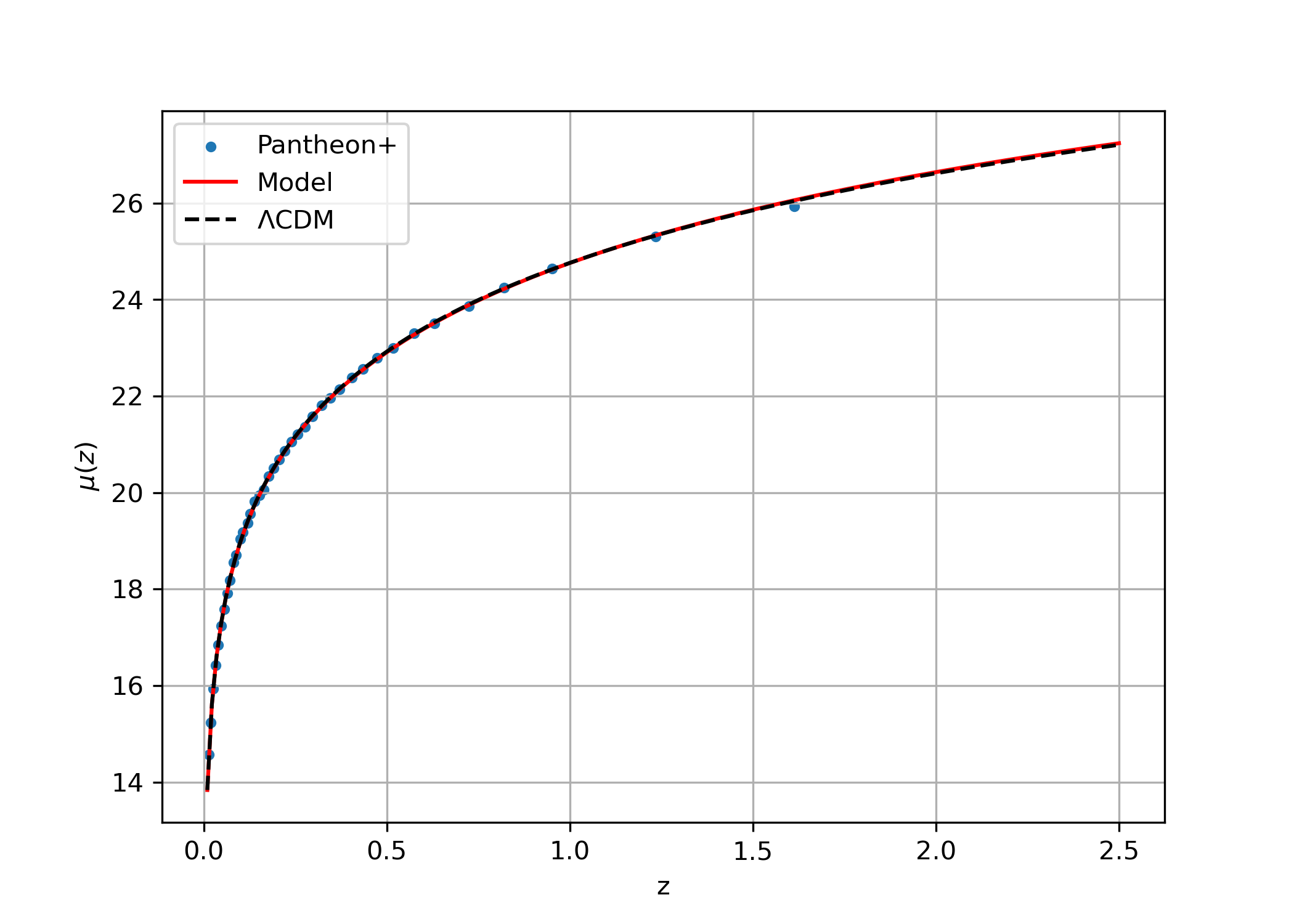}
	\caption{Comparison of the model with $\Lambda$CDM. Left panel: evolution of $H(z)$ with redshift. Right panel: variation of distance modulus $\mu(z)$.}
	\label{Hz_mu_plot}
\end{figure*}

In addition, Fig.~\ref{Hz_mu_plot} presents the comparison of the reconstructed model with the standard $\Lambda$CDM scenario. The left panel shows the evolution of the Hubble parameter $H(z)$ with redshift, while the right panel depicts the variation of the distance modulus $\mu(z)$. In both cases, the proposed model exhibits excellent agreement with the $\Lambda$CDM model, indicating its consistency with observational data.


\begin{table*}
	\centering
	\caption{Best-fit values of model parameters obtained from OHD, Pantheon+, and joint datasets.}
	\begin{tabular}{lcccc}
		\hline\hline
		Dataset & $\gamma$ & $\alpha$ & $H_0$ & $\chi^2_{\mathrm{min}}$ \\
		\hline
		OHD (77) & $0.5293^{+0.1756}_{-0.1658}$ & $0.4574^{+0.2235}_{-0.1770}$ & $63.71^{+3.18}_{-2.36}$ & 16.96 \\
		
		Pantheon+ (1701) & $0.2334^{+0.1872}_{-0.0722}$ & $1.0669^{+0.5348}_{-0.7015}$ & $69.62^{+6.85}_{-6.61}$ & 39.39 \\
		
		Joint & $0.2886^{+0.0395}_{-0.0347}$ & $0.8139^{+0.1007}_{-0.0954}$ & $68.62^{+1.35}_{-1.35}$ & 59.79 \\
		
		\hline\hline
	\end{tabular}
	\label{tab1}
\end{table*}


\section{Systematics and Limitations}

While the present analysis provides strong constraints on the model parameters, several sources of systematic uncertainties must be acknowledged. First, the OHD dataset, although model-independent, is affected by uncertainties in stellar population synthesis models used in the cosmic chronometer method. These uncertainties may introduce biases in the estimation of $H(z)$. Second, the Pantheon+ dataset relies on the calibration of Type Ia supernovae, which involves the nuisance parameter $M_B$. The degeneracy between $M_B$ and $H_0$ may influence the inferred cosmological parameters. Third, the assumption of Gaussian errors in the $\chi^2$ analysis may not fully capture potential non-Gaussian features in the data. Finally, the present study is restricted to background cosmology and does not include perturbation-level analyses such as structure formation or CMB anisotropies. Incorporating these effects in future work would provide a more stringent test of the model.

Despite these limitations, the consistency of the results with observational data and the agreement with $\Lambda$CDM behavior indicate that the proposed model remains a viable alternative for describing late-time cosmic acceleration.

\section{Physical and Kinematical Analysis}

In this section, we investigate the physical and kinematical behavior of the cosmological model using the best-fit values obtained from the joint observational analysis. The study of these parameters provides deeper insight into the expansion history, the nature of dark energy, and the overall dynamical evolution of the Universe. Before proceeding further, it is useful to establish some intermediate relations derived from the adopted parametrization of the Hubble parameter. For notational convenience, we define a dimensionless function $E(z) = 1 + \gamma \left((1+z)^{1+\alpha} - 1 \right)$, so that the Hubble parameter can be expressed in the compact form $H(z) = H_0 E(z)$. Differentiating with respect to the redshift $z$, we obtain $\frac{dH}{dz} = H_0 \gamma (1+\alpha)(1+z)^{\alpha}$. Further, using the standard relation between derivatives with respect to cosmic time $t$ and redshift $z$, $\dot{H} = \frac{dH}{dt} = -(1+z)H \frac{dH}{dz}$, we obtain $\dot{H} = -H_0^2 \gamma (1+\alpha)(1+z)^{1+\alpha} E(z)$. These relations will be used extensively in deriving various dynamical and kinematical quantities of the model.
\subsection{Deceleration Parameter}

The deceleration parameter $q(z)$ is a key kinematical quantity that characterizes the acceleration or deceleration of the cosmic expansion. It is observed as
\begin{equation}
	q(z) = -1 + \frac{\beta}{H_0} \left[ 1 + \gamma \left((1+z)^{1+\alpha} - 1 \right) \right]^{-1}.
\end{equation}

\begin{figure}[ht]
	\centering
	\includegraphics[scale=0.6]{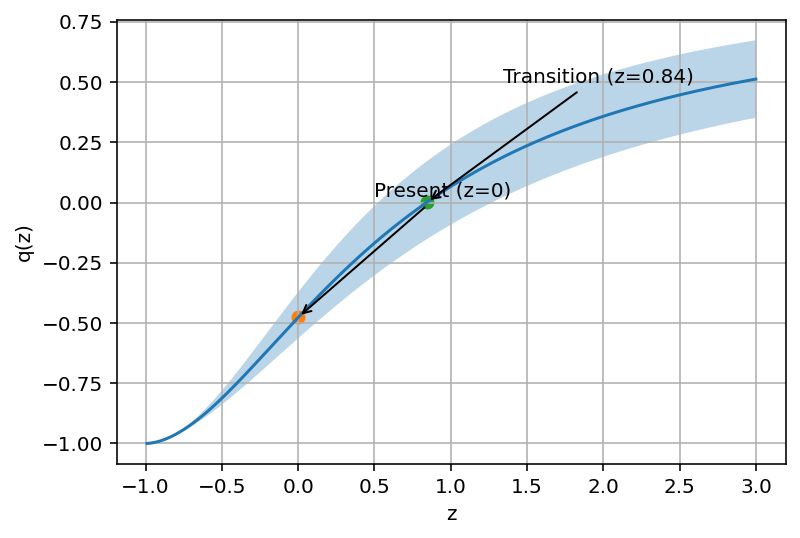}
	\caption{The behavior of $q$ versus redshift $z$ for the proposed model, using the best-fit values of the free parameters obtained from the joint analysis of OHD and Pantheon+ datasets.}.\label{q}
\end{figure}

\noindent In the present work, we analyze the dynamical evolution of the Universe through the behavior of the deceleration parameter $q(z)$ within the framework of $f(R,T)$ gravity. The variation of $q$ with redshift is illustrated in Fig.~\ref{q}, where the evolution is plotted using the best-fit values of the model parameters obtained from the joint analysis of the OHD and Pantheon+ datasets. 
It is evident from Fig.~\ref{q} that the deceleration parameter exhibits a clear transition from positive to negative values as the redshift decreases. This behavior reflects the well-established cosmological scenario in which the Universe evolves from an early decelerated phase to the present accelerated phase. The positive values of $q(z)$ at higher redshifts indicate a matter-dominated epoch, during which the expansion of the Universe was slowing down under the influence of gravitational attraction. As the Universe evolves, the deceleration parameter decreases and eventually crosses the zero line at a transition redshift $z_t \approx 0.84$. This point marks the onset of accelerated expansion, where the repulsive effect of dark energy begins to dominate over matter. At the present epoch ($z=0$), the model predicts a negative value of the deceleration parameter, $q_0 \approx -0.49$, which is consistent with current observational constraints from Type Ia Supernovae and cosmic microwave background measurements. Furthermore, in the asymptotic future limit ($z \rightarrow -1$), the deceleration parameter approaches $q \rightarrow -1$, indicating that the model asymptotically evolves towards a de Sitter phase. This behavior corresponds to an exponentially expanding Universe dominated by a constant dark energy component.

Overall, the obtained results are in close agreement with the predictions of the standard $\Lambda$CDM model, which also suggests a present value of the deceleration parameter around $q_0 \approx -0.5$ and a similar transition redshift. Hence, the proposed model successfully reproduces the observed expansion history of the Universe and provides a viable description consistent with both theoretical expectations and observational data.

\subsection{Statefinder Diagnostic}

To further characterize the model, we employ the statefinder parameters $(r,s)$ defined as
\begin{equation}
	r = \frac{\dddot{a}}{aH^3}=1+3\frac{\dot{H}}{H^{2}}+\frac{\ddot{H}}{H^{3}}, \quad s = \frac{r - 1}{3(q - 1/2)}.
\end{equation}

The trajectory of the model in the $(r,s)$ plane shows a clear evolution towards the $\Lambda$CDM fixed point $(r,s) = (1,0)$. This indicates that although the model deviates from $\Lambda$CDM at intermediate redshifts, it asymptotically approaches it at late times, consistent with observational expectations.
\begin{figure*}[ht]
	\centering
	\includegraphics[scale=0.4]{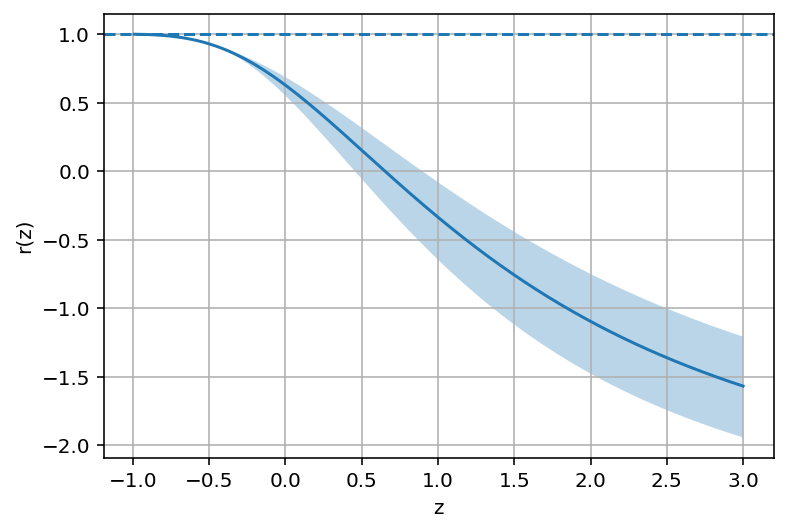}
	\includegraphics[scale=0.4]{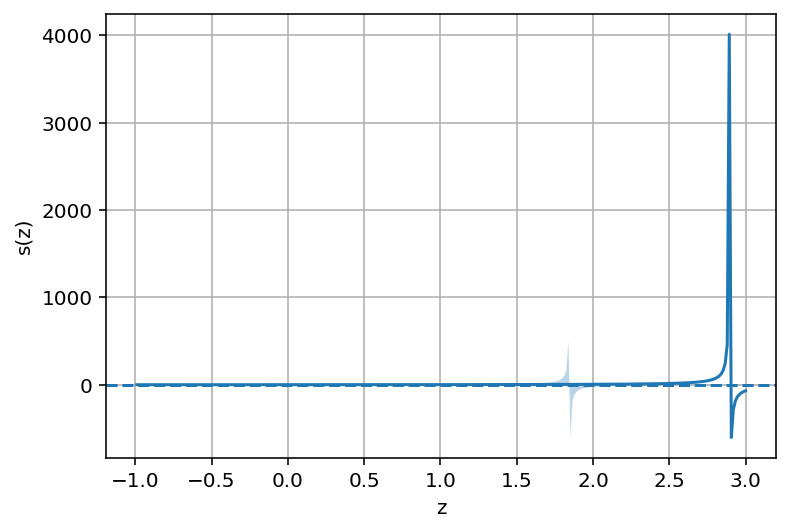}
	\includegraphics[scale=0.4]{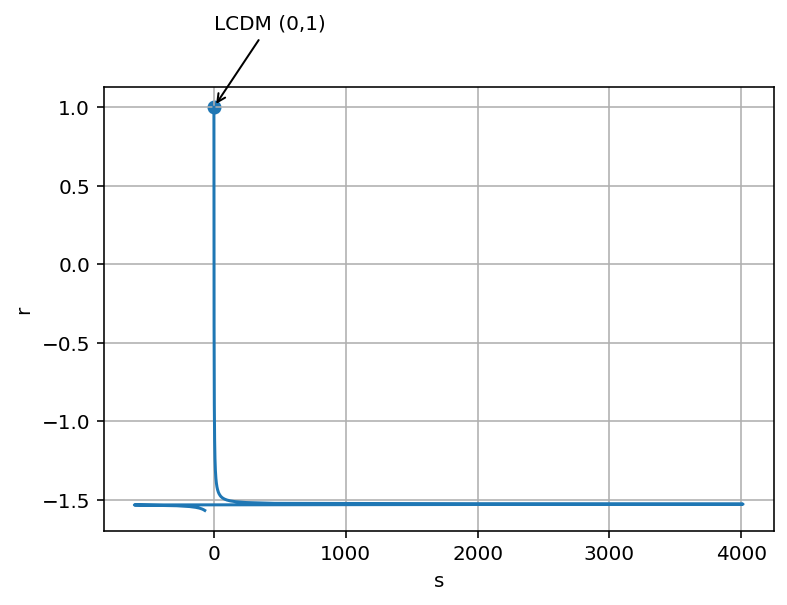}
	\caption{The behavior of $(r)$,  $(s)$ and $(r,s)$ - plane for the proposed model, using the best-fit values of the free parameters obtained from the joint analysis of OHD and Pantheon+ datasets.}\label{7}
\end{figure*}
\subsection{$O_m(z)$ Diagnostic}

The $O_m(z)$ diagnostic provides a useful tool to distinguish different dark energy models and is defined as
\begin{equation}
	O_m(z) = \frac{H^2(z)/H_0^2 - 1}{(1+z)^3 - 1}.
\end{equation}


\noindent The $O_m(z)$ diagnostic serves as an important geometrical tool to distinguish between different dark energy scenarios, such as quintessence ($\omega > -1$), phantom ($\omega < -1$), and the cosmological constant ($\omega = -1$). The behavior of the $O_m(z)$ function for the present model, constrained using the joint OHD and Pantheon+ datasets, is illustrated in Fig.~(\ref{8}).
\begin{figure}[H]
	\centering
	\includegraphics[scale=0.6]{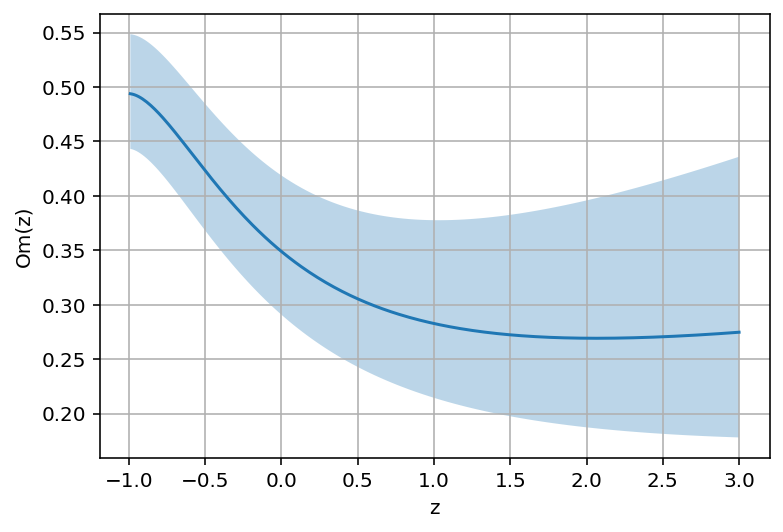}
	\caption{The behavior of $O_m(z)$ diagnostic versus $z$.}\label{8}
\end{figure}
For the standard flat $\Lambda$CDM model, the $O_m(z)$ parameter remains constant and is equal to the present matter density parameter $\Omega_{m0}$. Any deviation from constancy in $O_m(z)$ therefore indicates a departure from the $\Lambda$CDM paradigm and signals the presence of dynamical dark energy.

As shown in Fig.~(\ref{8}), the $O_m(z)$ curve corresponding to the present model exhibits a mild but noticeable redshift dependence, deviating slightly from a constant behavior. From the plot, the present-day value is estimated to be $O_m(0) \approx 0.35$, which lies close to the expected $\Lambda$CDM value, thereby indicating consistency with observational bounds. Furthermore, the overall increasing trend of $O_m(z)$ with redshift suggests a phantom-like behavior of the effective dark energy component. This interpretation follows from the known property that $O_m(z)$ increases with $z$ for $\omega < -1$ (phantom regime), while it decreases for $-1 < \omega < 0$ (quintessence regime) \cite{Sahni2008,Shafieloo2009}. This result is in good agreement with the behavior inferred from the effective equation of state parameter $\omega_{\rm eff}(z)$, as well as from the evolution of the deceleration parameter. In addition, similar redshift-dependent behavior of the $O_m(z)$ diagnostic has been reported in several modified gravity and dynamical dark energy models, particularly those allowing small deviations from the cosmological constant scenario \cite{Scolnic2018,Guo2005}. 

Therefore, the $O_m(z)$ analysis provides a consistent and complementary picture of the cosmological dynamics, indicating that the present $f(R,T)$ model effectively mimics a mildly phantom dark energy scenario while remaining compatible with current observational data.

\subsection{Equation of state parameter, stability, and energy conditions}

Using the reconstructed expressions of the dark energy pressure and
energy density obtained in Eqs.~(\ref{29}) and (\ref{30}), we now investigate the
equation of state parameter, classical stability, and energy conditions
of the model.\\
Observational evidence strongly supports a spatially flat Universe whose energy budget is dominated by dark energy (DE), contributing nearly $70\%$, while the remaining $\sim 30\%$ consists primarily of cold dark matter (CDM) and baryonic matter, with radiation playing a negligible role at late times. Although DE is widely accepted as the driving mechanism behind the present accelerated expansion, its underlying nature remains one of the central open problems in modern cosmology. A convenient way to characterize DE is through the equation of state (EoS) parameter, defined as $	\omega_s = \frac{p_s}{\rho_s}$, where $p_s$ and $\rho_s$ denote the pressure and energy density, respectively. Hence, the EoS parameter can be written as
\begin{widetext}
\begin{equation}
	\omega_s
	=
	\frac{
		\displaystyle
		\frac{2\lambda_1}{16\pi+3\lambda_2}
		\left(2\dot{H}+3H^2+\sigma^2\right)
		+
		\frac{\lambda_2}{16\pi+3\lambda_2}
		\left(3H^2-\sigma^2\right)
	}{
		\displaystyle
		-\rho_m
		+
		\frac{2\lambda_1}{16\pi+3\lambda_2}
		\left(3H^2-\sigma^2\right)
		+
		\frac{\lambda_2}{16\pi+3\lambda_2}
		\left(2\dot{H}+3H^2+\sigma^2\right)
	}.
	\label{eos}
\end{equation}
\end{widetext}
Various theoretical models have been proposed to explain the behavior of DE. In quintessence models, based on a dynamical scalar field, the EoS parameter lies in the range $-1 < \omega < -\frac{1}{3}$, allowing for accelerated expansion without invoking a cosmological constant. In contrast, phantom models correspond to $\omega < -1$, often associated with exotic properties and potential instabilities. Another class, known as quintom models, permits a dynamical transition across the cosmological constant boundary $\omega = -1$ \cite{Ratra1988,Sami2004,Elizalde2004,Nojiri2003,Armendariz2000,Padmanabhan2002,Khoury2004,Bento2002,Zarrouki2010}. Constraints on the present value of the EoS parameter have been obtained from a combination of observational datasets, including Type Ia Supernovae (SNIa), Cosmic Microwave Background (CMB), Baryon Acoustic Oscillations (BAO), and measurements of the Hubble parameter. The WMAP9 analysis suggests $\omega_0 = -1.084 \pm 0.063$ \cite{Hinshaw2023}, while the Planck mission reports $\omega_0 = -1.006 \pm 0.0451$ (2015) and $\omega_0 = -1.028 \pm 0.032$ (2018) \cite{Ade2015}.

\begin{figure}[ht]
	\centering
	\includegraphics[scale=0.6]{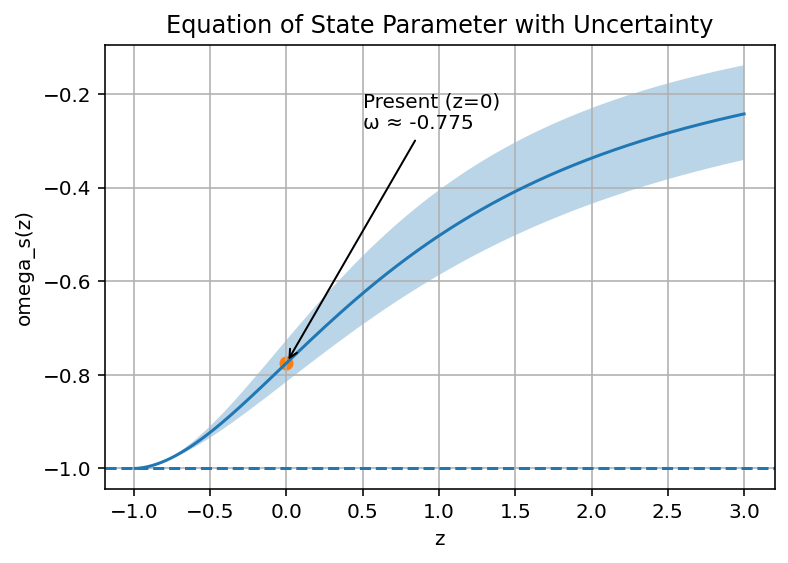}
	\caption{The behavior of $\omega_s$ versus redshift $z$ for the proposed model, using the best-fit values of the free parameters obtained from the joint analysis of OHD and Pantheon+ datasets.}.\label{w}
\end{figure}

In the present model, the EoS parameter $\omega(z)$ is reconstructed from the Hubble parameter through the $Om(z)$ diagnostic within the framework of $f(R,  T)$ gravity. The evolution of $\omega(z)$ provides important insight into the dynamical nature of the cosmic fluid. As illustrated in Fig.~(\ref{w}), the EoS parameter exhibits a clear redshift dependence. At earlier epochs ($z > 0$), $\omega(z)$ attains positive values, indicating a matter-dominated phase where pressure is negligible compared to energy density ($\omega \approx 0$). This behavior is consistent with the standard cosmological paradigm and confirms that the model successfully reproduces the expected early-time dynamics. At the present epoch ($z = 0$), the EoS parameter lies approximately in the range $-0.9 \lesssim \omega_0 \lesssim -0.8$, signifying a transition to an accelerated expansion phase. This interval falls within the quintessence regime, suggesting that the present acceleration may be driven by a dynamical dark energy component rather than a strictly constant cosmological term. As the redshift approaches $z \to -1$, corresponding to the far future, $\omega(z)$ gradually approaches $-1$, indicating that the model asymptotically mimics a $\Lambda$CDM-like behavior.
A notable feature of this evolution is that $\omega(z)$ remains greater than or equal to $-1$ throughout the cosmic history, thereby avoiding the phantom regime. Consequently, the weak energy condition is preserved, which strengthens the physical viability of the model. The smooth and monotonic evolution of $\omega(z)$ highlights the capability of the $f(R, T)$ gravity framework to describe a time-varying dark energy component without requiring exotic physics. Furthermore, the predicted present value of $\omega_0$ is in reasonable agreement with observational constraints. Although slightly higher than the central values reported by WMAP9 and Planck, it remains consistent within $1$--$2\sigma$ uncertainties and supports a quintessence-like scenario. Unlike quintom models, which allow crossing of the $\omega = -1$ boundary, the present model maintains a stable evolution approaching this limit without crossing it. Such deviations from a pure cosmological constant may be further tested with upcoming large-scale surveys such as DESI and Euclid, which are expected to provide tighter constraints on the dynamical properties of dark energy. Within current observational limits, however, the model predictions remain well within acceptable bounds.

Overall, the reconstructed behavior of the EoS parameter demonstrates that the $f(R, T)$ gravity model provides a consistent and physically acceptable description of cosmic evolution, successfully capturing the transition from a decelerated matter-dominated era to the present accelerated phase, while remaining compatible with both theoretical expectations and observational data.

\subsubsection{Stability analysis: squared sound speed}
To assess the physical viability of the reconstructed cosmological scenario, we examine its classical stability by analyzing the response of the cosmic fluid to small perturbations. A key quantity in this context is the squared sound speed, denoted by $\vartheta^2$, which characterizes the propagation of perturbations and is defined as the derivative of pressure with respect to energy density:
\begin{equation}
	\vartheta_s^2 = \frac{dp_s}{d\rho_s}
	= \frac{\dot{p}_s}{\dot{\rho}_s}.
\end{equation}

\begin{figure}[ht]
	\centering
	\includegraphics[scale=0.6]{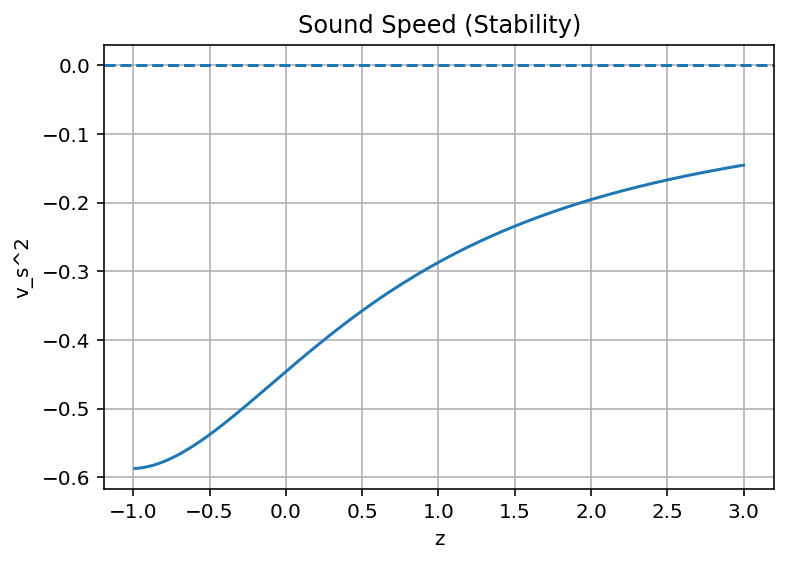}
	\caption{Evolution of the squared sound speed $\vartheta_s^2$ as a function of redshift $z$ for the proposed model, using the best-fit values of the free parameters obtained from the joint analysis of OHD and Pantheon+ datasets.}
	\label{v}
\end{figure}

Taking the time derivative of Eqs.~(29) and (30), 
Hence, the squared sound speed becomes
\begin{widetext}
\begin{equation}
\vartheta_s^2
	=
	\frac{
		\displaystyle
		\frac{2\lambda_1}{16\pi+3\lambda_2}
		\left(2\ddot{H}+6H\dot{H}+\dot{\sigma}^2\right)
		+
		\frac{\lambda_2}{16\pi+3\lambda_2}
		\left(6H\dot{H}-\dot{\sigma}^2\right)
	}{
		\displaystyle
		-\dot{\rho}_m
		+
		\frac{2\lambda_1}{16\pi+3\lambda_2}
		\left(6H\dot{H}-\dot{\sigma}^2\right)
		+
		\frac{\lambda_2}{16\pi+3\lambda_2}
		\left(2\ddot{H}+6H\dot{H}+\dot{\sigma}^2\right)
	}.
	\label{vs2}
\end{equation}
\end{widetext}
Within the framework of $f(R, T)$ gravity, the squared sound speed serves as a diagnostic tool to determine whether the model remains stable under linear perturbations. A positive value of $\vartheta_s^2$ indicates that perturbations propagate in a physically meaningful way, ensuring classical stability, whereas $\vartheta_s^2 < 0$ corresponds to an imaginary sound speed, signaling the growth of instabilities. The behavior of $\vartheta_s^2$ as a function of redshift is displayed in Fig.~(\ref{v}), covering the evolution from the early Universe ($z > 1$) to the future epoch ($z \to -1$). It is evident from the figure that $\vartheta_s^2$ remains negative throughout the considered redshift range, taking values approximately within the interval $-0.34 \lesssim \vartheta_s^2 \lesssim -0.28$. This persistent negativity implies that the model does not satisfy the classical stability condition and is therefore unstable against small perturbations.

Despite this limitation, the model successfully reproduces key features of the cosmic expansion history, including the transition from decelerated to accelerated expansion and a viable evolution of the equation of state parameter. It is worth noting that such instability is not uncommon in modified gravity theories and may arise due to the effective description of the cosmic fluid. In some cases, these issues can be mitigated by considering non-adiabatic perturbations, higher-order corrections, or quantum effects.

For comparison, standard quintessence models with canonical kinetic terms typically yield $\vartheta_s^2 = 1$, ensuring stability at the perturbative level. On the other hand, models with phantom-like behavior often exhibit $\vartheta_s^2 < 0$, similar to the trend observed here. This suggests that the present model may inherit certain features associated with such exotic components or may require further refinement to achieve complete physical consistency.

In summary, while the proposed model aligns well with observational constraints on the expansion dynamics, its classical instability, as indicated by the negative squared sound speed, highlights the need for additional investigation in order to establish a fully consistent cosmological framework.

\subsubsection{Energy conditions}

The physical acceptability of the model can be tested by examining the
energy conditions in terms of the reconstructed pressure and energy
density.

\paragraph{Null Energy Condition (NEC)}
\begin{equation}
	\rho_s + p_s \ge 0.
\end{equation}

Using Eqs.~(29) and (30), we find
\begin{equation}
	\rho_s + p_s
	=
	-\rho_m
	+
	\frac{2(\lambda_1+\lambda_2)}{16\pi+3\lambda_2}
	\left(2\dot{H}+6H^2\right).
\end{equation}

\paragraph{Weak Energy Condition (WEC)}
\begin{equation}
	\rho_s \ge 0,
	\qquad
	\rho_s + p_s \ge 0.
\end{equation}

\paragraph{Strong Energy Condition (SEC)}
\begin{equation}
	\rho_s + 3p_s \ge 0.
\end{equation}

Substituting Eqs.~(29) and (30), we obtain
\begin{small}
\begin{equation}
	\rho_s + 3p_s
	=
	-\rho_m
	+
	\frac{2(3\lambda_1+\lambda_2)}{16\pi+3\lambda_2}
	\left(2\dot{H}+3H^2\right)
	+
	\frac{2(3\lambda_1-\lambda_2)}{16\pi+3\lambda_2}\sigma^2.
\end{equation}
\end{small}

Violation of the strong energy condition at late times naturally
accounts for the accelerated expansion of the Universe.

\paragraph{Dominant Energy Condition (DEC)}
\begin{equation}
	\rho_s \ge |p_s|.
\end{equation}

Satisfaction of the null and weak energy conditions, together with the
violation of the strong energy condition, confirms the physical
viability of the present anisotropic $f(R,T)$ cosmological model.
\\

\begin{figure}[ht]
	\centering
	\includegraphics[scale=0.6]{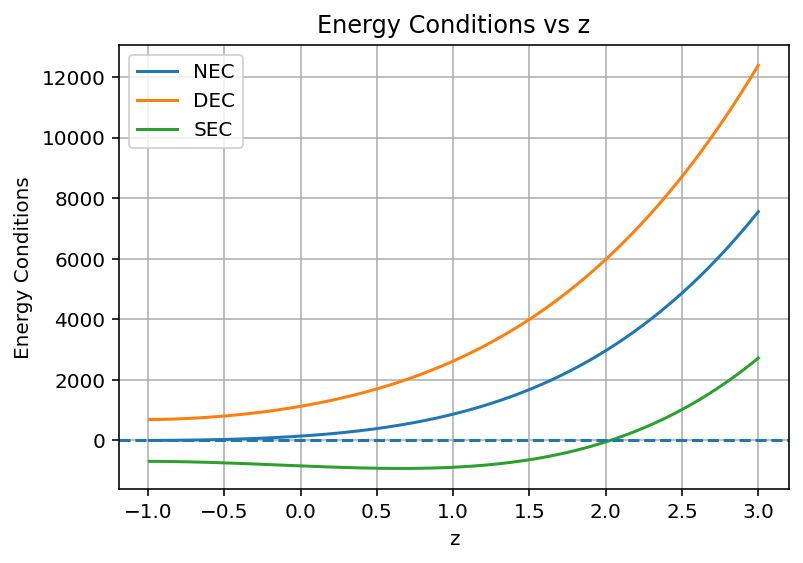}
	\caption{Evolution of the energy conditions as a function of redshift $z$ for the proposed model, using the best-fit values of the free parameters obtained from the joint analysis of OHD and Pantheon+ datasets.}
	\label{ec}
\end{figure}

The Null Energy Condition (NEC), which also ensures the Weak Energy Condition (WEC) for a perfect fluid, requires that $	\rho_s + p_s \geq 0$, where $\rho$ and $p$ represent the energy density and pressure of the effective cosmic fluid, respectively. This condition guarantees that the energy density measured by any null observer remains non-negative and is often regarded as a fundamental requirement for physically viable models. The evolution of $	\rho_s + p_s$ with redshift $z$ is illustrated in Fig.~(\ref{ec}). It is observed that $	\rho_s + p_s$ remains strictly positive over the entire range $z \in [-1, 3]$, indicating that both the NEC and WEC are satisfied throughout the cosmic evolution described by the model.
The fulfillment of this condition implies that the effective energy-momentum tensor does not admit exotic matter components such as those associated with wormholes or strongly phantom regimes, where $	\rho_s + p_s < 0$. Moreover, many well-established dark energy scenarios, including quintessence and the cosmological constant, also respect this condition. The present results are therefore consistent with both theoretical expectations and observational constraints, supporting the physical reliability of the model.

Next, we consider the Strong Energy Condition (SEC), which is expressed as $\rho_s + 3p_s \geq 0$. This condition reflects the requirement that gravity remains attractive under normal circumstances. However, in an accelerating Universe, particularly during the dark energy-dominated era, violation of the SEC is generally expected. The behavior of $\rho_s + 3p_s$ is shown in Fig.~(\ref{ec}), where it is evident that $\rho_s + 3p_s$ remains negative across the full redshift interval $z \in [-1, 2]$. This violation of the SEC is consistent with the observed late-time acceleration of the Universe and indicates the presence of a repulsive component in the cosmic dynamics.

Finally, the Dominant Energy Condition (DEC) imposes the requirement that $\rho_s \geq |p_s|$, which can be equivalently expressed as $\rho_s - p_s \geq 0$. This condition ensures that the energy density is non-negative and that the propagation of energy does not exceed the speed of light, thereby preserving causality. The evolution of $\rho_s - p_s$ with redshift is presented in Fig.~(\ref{ec}). The plot shows that $\rho_s - p_s$ remains positive and exhibits a monotonic increase throughout the range $z \in [-1, 3]$, confirming that the DEC is satisfied at all epochs. The preservation of the DEC indicates that the model avoids unphysical features such as negative energy densities or superluminal energy transport. This behavior is in agreement with standard cosmological models, including $\Lambda$CDM and scalar field-based dark energy scenarios. 

In summary, the combined analysis of energy conditions reveals that the present $f(R, T)$ cosmological model satisfies the NEC/WEC and DEC while violating the SEC, a behavior that is well-aligned with the requirements of an accelerating Universe. These results further support the physical plausibility and robustness of the model as a viable description of cosmic evolution.

\section{Conclusion}

In the present study, we have examined an anisotropic cosmological model within the framework of $f(R,T)$ gravity by incorporating interacting dark energy scenarios. The analysis has been carried out in a Bianchi type-I background, which provides a natural setting to investigate possible deviations from isotropy and their impact on the cosmic evolution. By adopting a reconstruction approach based on a variable deceleration parameter, we obtained an explicit redshift-dependent form of the Hubble parameter, enabling a direct comparison with observational data.

A detailed statistical analysis using observational Hubble data and  Pantheon+ supernova sample has performed to constrain the model parameters. The joint analysis significantly reduces uncertainties and yield parameter values that consistent with recent observational estimates. The reconstructed expansion history clearly demonstrate a smooth transition from a decelerated phase at an earlier epochs to  accelerated phase at late times. Furthermore, the model asymptotically evolve toward a de- Sitter phase, indicating that its capability to describe the long-term behavior of the Universe. The physical interpretation of the model has explored through various kinematical and geometrical diagnostics. The deceleration parameter confirms that the expected transition redshift and present-day acceleration. The statefinder analysis reveal that the model trajectory move towards the $\Lambda$CDM fixed point, suggested that the present framework can effectively mimic the standard cosmological model at late time. Similarly, the $O_m$ diagnostic indicates a mild deviation from a constant behavior, pointing toward a dynamical dark energy components rather than the strictly constant cosmological term.

The evolution of an equation of state parameter further supports this interpretation, as it describe a transition from a matter-dominated regime to a negative-pressure phase responsible for an accelerated expansion. Importantly, the model avoid crossing the phantom divide and remains within a physically acceptable range, which is consistent with current observational bounds. The analysis of energy conditions shows that the null, dominant and weak energy conditions are satisfied throughout cosmic evolution, while  strong energy condition is violated at late times, the feature that naturally accounts for the observed acceleration. The dominant energy condition is also preserved, ensuring causal consistency. From the perspective of anisotropy, the shear scalar decreases rapidly with the expansion of the Universe, which indicates that any initial anisotropy fades away at late times. This result ensures that the model effectively approaches an isotropic configuration consistent with present cosmological observations. However, the stability analysis based on the squared  speed of sound reveals that the model suffers from classical instability at the perturbative level. This issue suggested that additional mechanisms, such as non-adiabatic effects, higher-order corrections, or quantum contributions, may be required to achieve a fully consistent description.

In conclusion, the proposed $f(R,T)$ cosmological model with interacting  dark energy provides a coherent and observationally consistent framework to describe the late-time acceleration of the Universe. While it successfully captures key features of cosmic evolution and aligns closely with the $\Lambda$CDM paradigm, the presence of an instability highlighted the need for further refinement. Future investigations incorporating perturbation theory, structure formation, and high-precision cosmological observations may offers  a deeper investigation into the viability and completeness of such modified gravity models.

\section*{Authors contributions}
SHS: Writing-original draft, Software, Methodology, Investigation, Conceptualization. NM: Writing-review \& editing, Validation, Resources, Methodology. AP: Validation, Software, Investigation, Conceptualization. MZ: Validation, Software, Visualization, Supervision, Funding acquisition. 

\section*{Data availability}
No datasets were generated or analyzed during the current study.

\section*{Competing interests} The authors declare no competing interests.


\section*{Acknowledgments}
This research is funded by the Science Committee of the Ministry of Science and Higher Education of the Republic of Kazakhstan (Grant No. AP23483654). The IUCAA, Pune, India, is acknowledged by the author S. H. Shekh \& A. Pradhan for giving the facility through the Visiting Associateship programmes.

\begin{table*}[ht]
	\caption{Compilation of 77 observational Hubble parameter $H(z)$ measurements (in km\,s$^{-1}$Mpc$^{-1}$). Method I: Cosmic Chronometers, II: BAO (galaxy), III: BAO (Ly$\alpha$).}
	\centering
	\resizebox{\textwidth}{!}{
		\begin{tabular}{cccccccccc}
			\hline\hline
			S.No & $z$ & $H(z)$ & Method & Ref & S.No & $z$ & $H(z)$ & Method & Ref \\
			\hline
			
			1 & 0.00 & 69.1 $\pm$ 1.3 & I & \cite{Farooq} &
			40 & 0.68 & 93.9 $\pm$ 8.1 & I & \cite{Moresco} \\
			
			2 & 0.07 & 70.4 $\pm$ 20 & I & \cite{Zhang} &
			41 & 0.73 & 99.3 $\pm$ 7.1 & I & \cite{Blake} \\
			
			3 & 0.07 & 69.0 $\pm$ 19.6 & I & \cite{Zhang} &
			42 & 0.7812 & 105.0 $\pm$ 12 & I & \cite{Moresco} \\
			
			4 & 0.09 & 70.4 $\pm$ 12.2 & I & \cite{Simon} &
			43 & 0.781 & 107.1 $\pm$ 12.2 & I & \cite{Moresco} \\
			
			5 & 0.10 & 70.4 $\pm$ 12.2 & I & \cite{Zhang} &
			44 & 0.875 & 127.6 $\pm$ 17.3 & I & \cite{Moresco} \\
			6 & 0.12 & 68.6 $\pm$ 26.2 & I & \cite{Zhang} & 45 & 0.8754 & 125.0 $\pm$ 17 & I & \cite{Moresco} \\
			7 & 0.12 & 70.0 $\pm$ 26.7 & I & \cite{Farooq} & 46 & 0.88 & 91.8 $\pm$ 40.8 & I & \cite{Stern} \\
			8 & 0.17 & 83.0 $\pm$ 8 & I & \cite{Simon} & 47 & 0.880 & 90.0 $\pm$ 40 & I & \cite{Ratsimbazafy} \\
			9 & 0.17 & 84.7 $\pm$ 8.2 & I & \cite{Simon} & 48 & 0.90 & 69.0 $\pm$ 12 & I & \cite{Simon} \\
			10 & 0.179 & 76.5 $\pm$ 4 & I & \cite{Moresco} & 49 & 0.90 & 119.4 $\pm$ 23.4 & I & \cite{Simon} \\
			11 & 0.1791 & 75.0 $\pm$ 4 & I & \cite{Moresco} & 50 & 0.900 & 117.0 $\pm$ 23 & I & \cite{Simon} \\
			12 & 0.199 & 76.5 $\pm$ 5.1 & I & \cite{Moresco} & 51 & 1.037 & 157.2 $\pm$ 20.4 & I & \cite{Moresco} \\
			13 & 0.1993 & 75.0 $\pm$ 5 & I & \cite{Moresco} & 52 & 1.037 & 154.0 $\pm$ 20 & I & \cite{Moresco} \\
			14 & 0.20 & 72.9 $\pm$ 29.6 & I & \cite{Zhang} & 53 & 1.30 & 171.4 $\pm$ 17.3 & I & \cite{Simon} \\
			15 & 0.20 & 74.4 $\pm$ 30.2 & I & \cite{Zhang} & 54 & 1.300 & 168.0 $\pm$ 17 & I & \cite{Simon} \\
			16 & 0.24 & 81.5 $\pm$ 2.7 & II & \cite{Gaztanaga} & 55 & 1.363 & 160.0 $\pm$ 33.6 & a & \cite{Moresco2} \\
			17 & 0.27 & 78.6 $\pm$ 14.3 & I & \cite{Simon} & 56 & 1.363 & 163.3 $\pm$ 34.3 & I & \cite{Moresco2} \\
			18 & 0.28 & 88.8 $\pm$ 36.3 & I & \cite{Zhang} & 57 & 1.43 & 177.0 $\pm$ 18 & I & \cite{Simon} \\
			19 & 0.28 & 90.6 $\pm$ 37.3 & I & \cite{Farooq} & 58 & 1.43 & 180.6 $\pm$ 18.3 & I & \cite{Simon} \\
			20 & 0.35 & 84.4 $\pm$ 8.6 & II & \cite{Farooq} & 59 & 1.53 & 140.0 $\pm$ 14 & I & \cite{Simon} \\
			21 & 0.3519 & 83.0 $\pm$ 14 & I & \cite{Moresco} & 60 & 1.53 & 142.9 $\pm$ 14.2 & I & \cite{Simon} \\
			22 & 0.352 & 84.7 $\pm$ 14.3 & I & \cite{Moresco} & 61 & 1.75 & 202.0 $\pm$ 40 & I & \cite{Simon} \\
			23 & 0.38 & 81.5 $\pm$ 1.9 & II & \cite{Alam} & 62 & 1.75 & 206.1 $\pm$ 40.8 & I & \cite{Simon} \\
			24 & 0.3802 & 83.0 $\pm$ 13.5 & I & \cite{Moresco1} & 63 & 1.965 & 186.5 $\pm$ 50.4 & I & \cite{Moresco2} \\
			25 & 0.3802 & 84.7 $\pm$ 14.1 & I & \cite{Moresco1} & 64 & 1.965 & 190.3 $\pm$ 51.4 & I & \cite{Moresco2} \\
			26 & 0.40 & 95.0 $\pm$ 17 & I & \cite{Simon} & 65 & 2.30 & 228.0 $\pm$ 8.1 & III & \cite{Delubac} \\
			27 & 0.40 & 96.9 $\pm$ 17.3 & I & \cite{Simon} & 66 & 2.34 & 226.5 $\pm$ 7.1 & III & \cite{Delubac} \\
			28 & 0.4004 & 77.0 $\pm$ 10.2 & I & \cite{Moresco1} & 67 & 2.36 & 230.6 $\pm$ 8.2 & III & \cite{Font-Ribera} \\
			29 & 0.4004 & 78.6 $\pm$ 10.4 & I & \cite{Moresco1} & 68 & 0.4247 & 87.1 $\pm$ 11.2 & I & \cite{Moresco1} \\
			30 & 0.4247 & 87.1 $\pm$ 11.2 & I & \cite{Moresco1} & 69 & 0.4247 & 88.9 $\pm$ 11.4 & I & \cite{Moresco1} \\
			31 & 0.4247 & 88.9 $\pm$ 11.4 & I & \cite{Moresco1} & 70 & 0.4497 & 92.8 $\pm$ 12.9 & I & \cite{Moresco1} \\
			32 & 0.43 & 88.3 $\pm$ 3.8 & I & \cite{Farooq} & 71 & 0.4497 & 94.7 $\pm$ 13.1 & I & \cite{Moresco1} \\
			33 & 0.44 & 84.3 $\pm$ 7.9 & I & \cite{Blake} & 72 & 0.470 & 89.0 $\pm$ 34.0 & I & \cite{Ratsimbazafy} \\
			34 & 0.4497 & 92.8 $\pm$ 12.9 & I & \cite{Moresco1} & 73 & 0.47 & 90.8 $\pm$ 50.6 & I & \cite{Ratsimbazafy} \\
			35 & 0.4497 & 94.7 $\pm$ 13.1 & I & \cite{Moresco1} & 74 & 0.4783 & 80.0 $\pm$ 99.0 & I & \cite{Moresco1} \\
			36 & 0.470 & 89.0 $\pm$ 34.0 & I & \cite{Ratsimbazafy} & 75 & 0.4783 & 82.5 $\pm$ 9.2 & I & \cite{Moresco1} \\
			37 & 0.47 & 90.8 $\pm$ 50.6 & I & \cite{Ratsimbazafy} & 76 & 0.48 & 99.0 $\pm$ 63.2 & I & \cite{Ratsimbazafy} \\
			38 & 0.4783 & 80.0 $\pm$ 99.0 & I & \cite{Moresco1} & 77 & 0.64 & 98.82 $\pm$ 2.98 & II & \cite{Wang1} \\
			39 & 0.4783 & 82.5 $\pm$ 9.2 & a & \cite{Moresco1} &  &  &  &  &  \\
			\hline\hline
		\end{tabular}\label{tab}
	}
\end{table*}


\begin{thebibliography}{99}
			
			\bibitem{Riess1998}
			A.~G.~Riess et al., Astron. J. \textbf{116}, 1009 (1998).
			
			\bibitem{Perlmutter1999}
			S.~Perlmutter et al., Astrophys. J. \textbf{517}, 565 (1999).
			
			\bibitem{Spergel2003}
			D.~N.~Spergel et al., Astrophys. J. Suppl. \textbf{148}, 175 (2003).
			
			\bibitem{Planck2018} N Aghanim et al., 
			 Astron. Astrophys. \textbf{641}, A6 (2020).
			
			\bibitem{Eisenstein2005}
			D.~J.~Eisenstein et al., Astrophys. J. \textbf{633}, 560 (2005).
			
			\bibitem{Weinberg1989}
			S.~Weinberg, Rev. Mod. Phys. \textbf{61}, 1 (1989).
			
			\bibitem{Carroll2001}
			S.~M.~Carroll, Living Rev. Relativ. \textbf{4}, 1 (2001).
			
			\bibitem{Ratra1988}
			B.~Ratra and P.~J.~E.~Peebles, Phys. Rev. D \textbf{37}, 3406 (1988).
			
			\bibitem{Caldwell1998}
			R.~R.~Caldwell, R.~Dave and P.~J.~Steinhardt, Phys. Rev. Lett. \textbf{80}, 1582 (1998).
			
			\bibitem{Caldwell2002}
			R.~R.~Caldwell, Phys. Lett. B \textbf{545}, 23 (2002).
			
			\bibitem{Guo2005}
			Z.~K.~Guo et al., Phys. Lett. B \textbf{608}, 177 (2005).
			
			\bibitem{Armendariz2001}
			C.~Armendariz-Picon et al., Phys. Rev. D \textbf{63}, 103510 (2001).
			
			\bibitem{Nojiri2006}
			S.~Nojiri and S.~D.~Odintsov, Phys. Rev. D \textbf{74}, 086005 (2006).
			
			\bibitem{Capozziello2006}
			S.~Capozziello et al., Phys. Lett. B \textbf{639}, 135 (2006).
			
			\bibitem{Nojiri2005}
			S.~Nojiri and S.~D.~Odintsov, Phys. Lett. B \textbf{631}, 1 (2005).
			
			\bibitem{Harko2011}
			T.~Harko et al., Phys. Rev. D \textbf{84}, 024020 (2011).
						
				\bibitem{Singh2025}
				K.~Singh, A.~K.~Yadav and V.~K.~Bhardwaj, 
				Astrophys.\ Space Sci.\ (2025).
				
				\bibitem{Kumar2025}
				R.~Kumar, S.~K.~Tripathy and P.~Sahoo, 
				Eur.\ Phys.\ J.\ Plus (2025).
				
				\bibitem{Hernandez2024}
				A.~Hernandez-Almada, M.~A.~Garcia-Aspeitia and J.~Magaña, 
				Eur.\ Phys.\ J.\ C (2024).
				
				\bibitem{Sharif2025}
				M.~Sharif and A.~Waseem, 
				Astropart.\ Phys.\ (2025).
				
				\bibitem{Bahamonde2024}
				S.~Bahamonde, et al., 
				Eur.\ Phys.\ J.\ C (2024).
				
				\bibitem{Moraes2024}
				P.~H.~R.~S.~Moraes, et al.,
				``Wormhole solutions in $f(R,T)$ gravity,''
				Universe \textbf{7}, 43 (2024).
				
			
			
			
			
			\bibitem{Shabani2014}
			H.~Shabani and M.~Farhoudi, Phys. Rev. D \textbf{90}, 044031 (2014).
			
			\bibitem{Alvarenga2013}
			F.~G.~Alvarenga et al., JCAP \textbf{1309}, 007 (2013).
			
			\bibitem{tHooft1993}
			G.~'t Hooft, arXiv:gr-qc/9310026.
			
			\bibitem{Susskind1995}
			L.~Susskind, J. Math. Phys. \textbf{36}, 6377 (1995).
			
			\bibitem{Li2004}
			M.~Li, Phys. Lett. B \textbf{603}, 1 (2004).
			
			\bibitem{Tavayef2018}
			M.~Tavayef et al., Phys. Lett. B \textbf{781}, 195 (2018).
			
			\bibitem{Moradpour2018}
			H.~Moradpour et al., Eur. Phys. J. C \textbf{78}, 829 (2018).
			
			\bibitem{Jahromi2018}
			A.~S.~Jahromi et al., Phys. Lett. B \textbf{780}, 21 (2018).
			
			\bibitem{Ellis1969}
			G.~F.~R.~Ellis, J. Math. Phys. \textbf{10}, 1451 (1969).
			
			\bibitem{Collins1973}
			C.~B.~Collins and S.~W.~Hawking, Astrophys. J. \textbf{180}, 317 (1973).
			
			\bibitem{Berman1983}
		M. S. Berman,
		A special law of variation for Hubble's parameter,
		\textit{Nuovo Cimento B} \textbf{74}, 182 (1983).
		
		\bibitem{Abdussattar1997}
		Abdussattar and R. G. Vishwakarma, 
		\textit{Classical and Quantum Gravity} \textbf{14}, 945 (1997).
		
		\bibitem{Pradhan2012}
		A. Pradhan and H. Amirhashchi, 
		\textit{Modern Physics Letters A} \textbf{26}, 2261 (2011).
		
		\bibitem{Sahni2003}
		V. Sahni, T. D. Saini, A. A. Starobinsky, and U. Alam, 
		\textit{JETP Letters} \textbf{77}, 201 (2003).
		
		\bibitem{Shafieloo2006}
		A. Shafieloo, 
		\textit{Mon. Not. R. Astron. Soc.} \textbf{380}, 1573 (2007).
		
		
		
		\bibitem{Moresco2020} M. Moresco et al., \textit{Astrophys. J.} \textbf{898}, 82 (2020).
		\bibitem{Farooq} O. Farooq, et al., Astrophys. J., \textbf{835}, 026 (2017). 
		\bibitem{Moresco} M. Moresco, A. Cimatti, R. Jimenez et al., J. Cosm. Astropart. Phys., \textbf{1208}, 006 (2012).
		\bibitem{Zhang} M.-J. Zhang and J. -Q. Xia, J. Cosm. Astropart. Phys., \textbf{12}, 005 (2016).
		\bibitem{Blake} C. Blake, et al.,  Mon. Not. Roy. Astron. Soc., \textbf{425}, 405 (2012). 
		\bibitem{Simon} J. Simon, L. Verde and R. Jimenez, Phys. Rev. D \textbf{71}, 123001 (2005).
		\bibitem{Stern} D. Stern, R. Jimenez, L. Verde, et al. J. Cosm. Astropart. Phys., \textbf{1002}, 008 (2010).
		\bibitem{Ratsimbazafy} A. L. Ratsimbazafy, S. I. Loubser,  S. M. Crawford, et al.,  Mon. Not. Roy. Astron. Soc., \textbf{467}, 3239 (2017). 
		\bibitem{Gaztanaga} Gaztanaga, E., et al. Mon. Not. Roy. Astron. Soc., \textbf{399}, 1663 (2009).
		\bibitem{Moresco2} Moresco, M., Mon. Not. Roy. Astron. Soc., \textbf{450}, L16 (2015).
		\bibitem{Alam} S. Alam, M. Ata, S. Bailey et al., Mon. Not. Roy. Astron. Soc. \textbf{470}, 2617 (2017).
		\bibitem{Moresco1} M. Moresco, L. Pozzetti, A. Cimatti et al., J. Cosm. Astropart. Phys., \textbf{1605}, 014 (2016). 
		\bibitem{Delubac} T. Delubac, J. F. Bautista, N. G. Busca, et al., Astron. Astrophys., \textbf{574}, A59 (2015).
		\bibitem{Font-Ribera} A. Font-Ribera, D. Kirkby, N. Busca, et al., J. Cosmol. Astropart. Phys., \textbf{05}, 027 (2014).
		\bibitem{Wang1} Y. Wang, et al.,  Mon. Not. Roy. Astron. Soc., \textbf{469}, 3762 (2017). 
		\bibitem{Sahni2008}
		V.Sahni, A.Shafieloo, and A.A.Starobinsky,
		``Two new diagnostics of dark energy,''
		Phys.\ Rev.\ D {\bf 78}, 103502 (2008).
		\bibitem{Shafieloo2009} A. Shafieloo et al., Is cosmic acceleration slowing down?, Phys. Rev. D, (2009), 80, (083528).
		
		\bibitem{Scolnic2018} D. M. Scolnic et al., \textit{Astrophys. J.} \textbf{859}, 101 (2018).
		
		
		\bibitem{Caldwell2002} R. R. Caldwell, \textit{Phys. Lett. B} \textbf{545}, 23 (2002).
		\bibitem{Sami2004} M. Sami and S. Toporensky, \textit{Mod. Phys. Lett. A} \textbf{19}, 1509 (2004).
		\bibitem{Elizalde2004} E. Elizalde, S. Nojiri and S. D. Odintsov, \textit{Phys. Rev. D} \textbf{70}, 043539 (2004).
		\bibitem{Nojiri2003} S. Nojiri and S. D. Odintsov, \textit{Phys. Lett. B} \textbf{562}, 147 (2003).
		\bibitem{Armendariz2000} C. Armendariz-Picon, V. Mukhanov and P. J. Steinhardt, \textit{Phys. Rev. Lett.} \textbf{85}, 4438 (2000).
		\bibitem{Padmanabhan2002} T. Padmanabhan, \textit{Phys. Rev. D} \textbf{66}, 021301 (2002).
		\bibitem{Khoury2004} J. Khoury and A. Weltman, \textit{Phys. Rev. D} \textbf{69}, 044026 (2004).
		\bibitem{Bento2002} M. C. Bento, O. Bertolami and A. A. Sen, \textit{Phys. Rev. D} \textbf{66}, 043507 (2002).
		\bibitem{Zarrouki2010} R. Zarrouki and M. Bennai, \textit{Int. J. Mod. Phys. A} \textbf{25}, 2507 (2010).
		\bibitem{Hinshaw2023} G. Hinshaw et al., \textit{Astrophys. J. Suppl. Ser.} \textbf{208}, 19 (2023).
		
		\bibitem{Ade2015} P. A. R. Ade et al., \textit{Astron. Astrophys.} \textbf{594}, A13 (2015).
		
		
		\end{thebibliography}
\end{document}